# TWIN SUNS IN AUSTRALIAN ABORIGINAL TRADITIONS

Duane W. Hamacher

*Monash Indigenous Studies Centre, Monash University, Clayton, Victoria 3800, Australia, and the Centre for Astrophysics, University of Southern Queensland, Toowoomba, QLD, 4350, Australia.*
Email: duane.hamacher@gmail.com

and

Rubina R. Visuvanathan

*World Vision International, Plaza Hamodal, Lot 15, Jalan 13/2, Section 13, Petaling Jaya, 46200 Selangor, Malaysia.*
Email: rubeca@gmail.com

**Abstract:** The oral traditions of Aboriginal cultures across Australia contain references to the presence of multiple Suns in the sky at the same time. Explanations of this have been largely regarded as symbolic or mythological, rather than observations of natural phenomena. In this paper, we examine oral traditions describing multiple Suns and analyse interpretations that could explain them. Our analysis of the oral traditions concludes that descriptions of multiple Suns fall into two main categories: one describing the changes in the path of the Sun throughout the year, and the other describing observations of parhelia, an atmospheric phenomenon known as 'Sun dogs' that creates an optical illusion of multiple Suns in the sky at once. This analysis shows how Aboriginal people pay close attention to natural phenomena, assign them social meaning, and incorporate them into oral tradition.

**Keywords:**  Indigenous Australians, Cultural Astronomy, ethnoastronomy, ethnometeorology, parhelia

## 1    INTRODUCTION

Oral traditions are highly complex, containing multiple layers of symbolism, allegory, and knowledge (Kelly, 2015) – the latter of which include descriptions and explanations of various natural processes (Hamacher, 2012). These traditions are often transmitted through narratives around storylines featuring humans, animals, or supernatural beings performing feats of heroism and strength that describe the formation of the world and inform sacred law. These traditional and cultural practices help record complex information to memory and pass it down to successive generations.

The study of the astronomical knowledge and traditions of Aboriginal and Torres Strait Islander people, using the theoretical frameworks and methodologies of cultural astronomy, has revealed a wealth of information about the ways in which Indigenous peoples conceptualize and utilize the sky (Clarke, 2009; Norris, 2016). Sky knowledge is intimately linked with the land and sea, and informs sacred law, kinship, and ceremony. This is encoded in a dynamic and holistic knowledge system based upon the observations and experiences of people, which was passed down the generations through oral tradition and material culture.

Descriptions and explanations of transient celestial phenomena are commonplace in Indigenous Knowledge Systems across the globe and Australia, including comets (Hamacher and Norris, 2011a), variable stars (Hamacher and Frew, 2010; Leaman et al. 2014; Hamacher, 2014; Hamacher 2018), and eclipses (Hamacher and Norris, 2011b), but also includes atmospheric phenomena, such as meteors (Hamacher and Norris, 2010) and aurorae (Hamacher, 2013). Although the weather and climate knowledge of Indigenous Australians is being documented (Clarke, 2009; Green, et al., 2010: 223), many types of rare atmospheric phenomena remain poorly understood from the Indigenous Knowledge perspective.

Some Aboriginal oral traditions narrate the simultaneous presence of multiple Suns in the sky. Similar narratives are found in the oral traditions and material culture of Indigenous cultures of the world (e.g. Eells 1889: 680; Gallagher, 2001: 510; Gusinde 1937: 1145-1146; James and Van der Sluijs, 2016: 70; Kováč, 2017: 19; MacDonald, 1998: 158; Needham and Ronan 1981: 229; Sassen, 1994: 4757), but descriptions of multiple Suns in Australian Aboriginal oral traditions have not been studied in any detail.

Many questions are posed when trying to address this topic. Are these descriptions of a witnessed phenomenon, or are they symbolic in meaning and interpretation? How are these types of phenomena conceptualised, and what significance to Aboriginal people place on them? We aim to address these questions through a targeted study of oral traditions describing multiple Suns in the sky. As a first step, we hypothesize that some of the stories of multiple Suns in the sky represent cultural explanations of atmospheric halo phenomena. To test this, we examine oral traditions from across Australia that tell of multiple Suns, as well as published literature that attempts to explain them. For example, some scholars suggest the appearance of multiple Suns may represent supernovae (Iqbal *et al.*, 2009) or a literal interpretation of a stellar companion (Andrews, 2004), while others suggest they are allegorical or symbolic descriptions (Haynes, 1990). We systematically test these claims and examine natural phenomena that best fit the narrative in the oral traditions.

This paper is part of an ongoing series that examines oral traditions related to astronomical and atmospheric phenomena, using a combination of archival and ethnographic studies. This paper is a literature and archival study to test a particular hypothesis. Current ethnographic studies with Aboriginal and Torres Strait Islander communities will address the questions raised, and examine how these phenomena are conceptualised and understood in a social context.

## 2     TWIN SUN TRADITIONS

A survey of published literature and archival materials regarding any mention of multiple Suns in the sky in Aboriginal oral traditions resulted in six oral traditions of multiple or phantom Suns. These traditions originate from Elcho Island, Yirrkalla, and Oenpelli in Arnhem Land (Northern Territory), the Tiwi Islands (Northern Territory), Cape Bedford (Queensland), and the Kimberley region (Western Australia). In part, this northern bias is likely the product of the existence of more extensive records of Indigenous astronomical traditions being recorded in oral and material tradition.

### 2.1     Arnhem Land, NT

Various Yolngu oral traditions depict the Sun as a woman with a daughter or daughters. On Elcho Island, the people portray the Sun as a woman and the Moon as her husband. They have several children, who are also Suns. They live below the horizon in the east. The Sun woman does not bring them with her as she crosses the sky because if she did, she would inadvertently kill the Dhuwa people (people of a different moiety) (Warner, 1937: 538). After she sets in the west, she hurries back to the east to care for her children before journeying across the sky again. The tradition holds an important social lesson about the ramifications of domestic abuse: Warner notes that the people believed that "If they are hit by a man or disturbed in any way, the rain won't come and the wells will dry up." The traditions also note the relative motions of the Sun and Moon, explaining that solar eclipses are the two making love (*ibid*.).

An oral tradition from the Yirrkalla community in northeastern Arnhem Land narrates the presence of two Suns in the sky; a mother and her daughter (Mountford 1956: 502; Roberts and Roberts 1989: 54). The Sun-woman, Walu, and her daughter, Bara, rise together in the eastern horizon to begin their journey across the sky. But Walu feared that the combined heat of both her and her daughter would burn the people and set the land on fire, so she sent the daughter away to below the horizon. This story is featured in bark paintings that illustrate the narrative (Fig. 1).

The anthropologists Ronald and Catherine Berndt recorded a tradition about two Suns from Balgo, Western Australia (Berndt and Bernt, 1989: 370-371), describing the interaction between the Moon man (Wadi Yagan) and two Sun women (Djindu). The story describes Wadi Yagan hunting possum but having no luck. In his anger, he attacked the dancing Djindu and killed them, but they came back to life and spend all day dancing and all night resting.

Berndt (1952: 46) recorded a similar version of the oral tradition in which the Sun has two daughters that once accompanied her across the sky. She sent the daughters back because three Suns in the sky would be too hot for the inhabitants of Earth. In 1953, Richard Waterman related to Mountford (1956: 502) another tradition from Yirrkalla that speaks about the occasional presence of two Suns in the sky: a husband and wife.

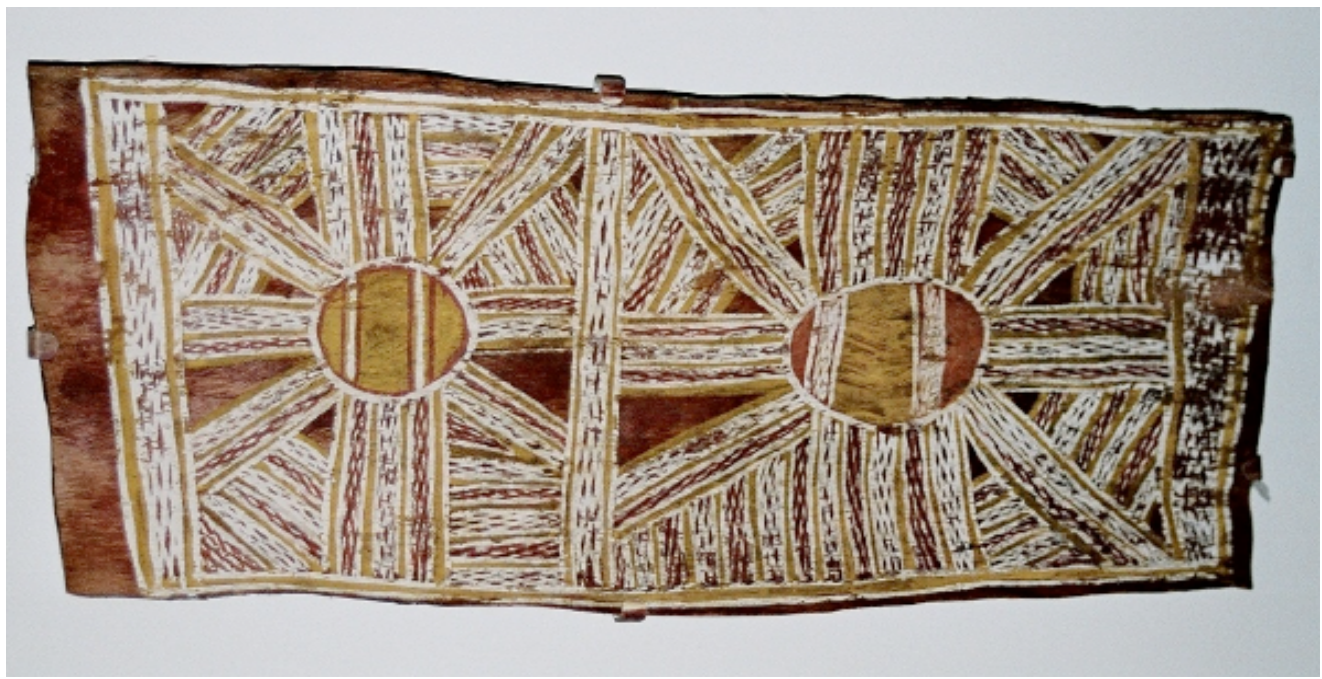

**Figure 1:** A bark painting from Yirrkalla showing the story of Walu the Sun Woman. The radiating lines represent the rays of the Sun as it rises and sets. The two bars across the Sun represent red feathered string ornaments that she always wears around her waist and forehead (Mountford 1956: 497).

## 2.2  Tiwi Islands, NT

In Tiwi traditions of Bathurst and Melville Islands, north of Darwin, there are two sister Suns: Wuriupranala and Murupiangkala (Mountford, 1958: 25, 41, 173). They are the daughters of the creation ancestor, Mudungkala, who is described as an old, blind woman. She emerged from the underworld to care for her children, creating the land in the process. Her children comprised two daughters named Wuriupranala and Murupiangkala and a son named Purukupali. There was no light and no heat at this time. Two ancestor men (who represented the Wedge-Tailed Eagle and the Fork-Tailed Kite) rubbed two sticks together, accidentally creating fire. Purukupali realised the light (milaijuka) and heat (ikwani) were beneficial, and lit a bark torch, which he gave to his sister, Wuriupranala. It was her responsibility to keep it lit. The Moon man, Tjapara (whose face is scarred from a fight), was given a smaller torch. At the start of every day, the Sun woman covers herself in red ochre and lights the bark torch of the Sun. Some of the ochre falls off of her and creates the red glow in the morning sky. As she sets in the west, she again powders her body with ochre, extinguishes her torch and goes into the underworld, finding her way back east using the glowing embers of the torch. The Moon man follows a similar path in the sky.

Mountford (1958: 173) describes a tradition, and the only case, in which a second Sun woman is identified: Murupiangkala, the sister of Wuriupranala. He describes how this story is encoded in a bark painting (Figure 2):

> *"Wuriupranala and Murupiangkala. One of the Sun-women, Wuriupranala, is shown at **f** (upper edge) the circle of radiating lines, **g**, symbolizing the flames of her bark torch. At **h**, on the bottom of the painting, is the other Sun-woman, Murupiangkala with her bark torches. The Moon-man, Tjapara, is at **b**, his head at **e**, his shoulders at **c**, and his genitals at **d**. The oval at **k** (upper right), is a mountain in the east from the top of which Wuriupranala or Murupiangkala alternately set out on their daily journey across the sky. The white oval at **l** (upper left) indicates the mountain on the western horizon where the Sun-women rest for a while before they travel to the Kumpinulu lagoon, and prepare and eat their meal before returning to the east via the underground road, **m** (upper left). The designs on the lower edge of the painting deal with the journey of the Sun-woman, Murupiangkala; the white design **o**, is the mountain in the east from which she sets out on her journey across the sky; **p** represents the mountain in the west where she finishes her journey."*

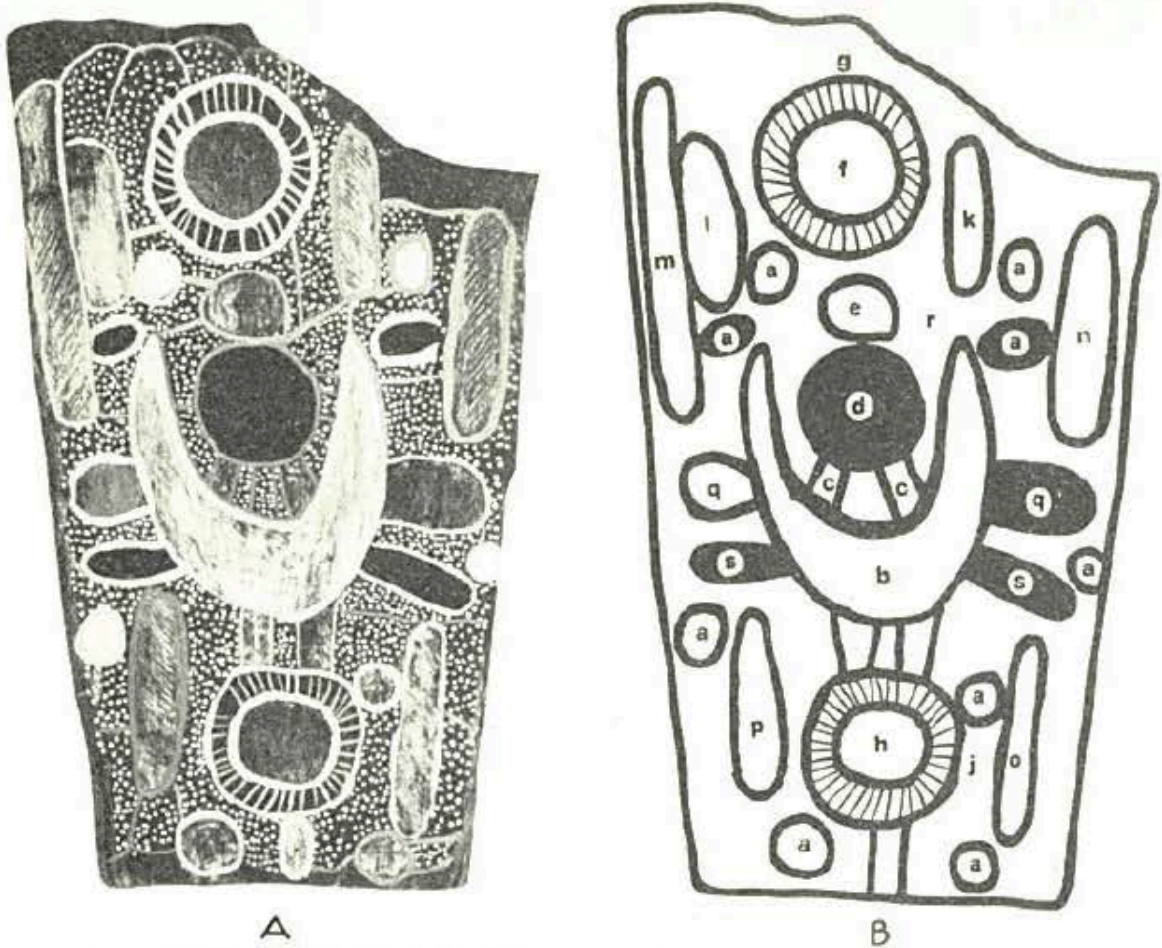

**Figure 2:** A Tiwi bark painting describing the two Sun women (after Mountford, 1958: 173).

The unidentified artist who made the painting said the Sun women travel alternately across the sky, though Mountford says this is "was not the dominant belief."

### 2.3   Cape Bedford, QLD

Roth (1901: 7) recorded an oral tradition at the Cape Bedford Lutheran Mission in northern Queensland at the end of the 19[th] century. The Aboriginal people (possibly the Guugu Yimidhirr) said that the Moon is a man and the Sun is his wife. The Moon feeds on fish until he becomes fully gorged (Full Moon), then wastes away until he starves and dies (New Moon), before going fishing again (waxing Moon). Although the Sun is referred to in the singular form, the Sun "Wife" actually comprises two sisters. The Sun is always moving westward as she hunts for green ants (also called weaver ants, of the *Oecophylla* genus), whose larvae are a highly sought food source. The Aboriginal people say that the elder sister-Sun visits them in the cold season (winter) while the younger sister visits them in the hot season (summer). In this case, the tradition never declares the sister Suns are visible in the sky at the same time, but rather describes the two general pathways of the Sun at different times of the year.

### 2.4   The Kimberley, WA

David Mowaljarlai – a Worrorra man from the Mowanjum community near Derby in the Kimberley region of Western Australia – describes a relationship between the Mother and Daughter Suns (Utemorrah et al., 1980: 102–103). In the tradition, the Mother Sun caused a great drought that dried up all the creeks and rivers, burnt the grass, and killed the animals. The Suns thought about how they could prevent this from happening again. The two Sun women lived in a world to the east, beyond the land and sea. They lived in three logs, which were smooth and slippery. These logs were gateways to our world, and the mother Sun grew too big and became stuck. Her daughter could easily move in and out of the log and because she shone, decided to go across the sky in her mother's place. As she journeyed across the sky, she trod on a snake, which bit her on the thigh.

This made her very ill and she died at dusk. Her mother nursed her back to life and explained to her daughter that she must relive this journey every day. This is so their combined heat does not burn the land and cause another drought.

The story and its variants (e.g. Clendon, 1999: 316) contain further details, but do not assert both Suns were in the sky at the same time. The story provides an explanation regarding the daily motion of the Sun. The story suggests that at one time, both Suns were present in the sky at once, which caused the drought from their combined heat. This is similar to the story from Oenpelli.

## 3   Discussion

An analysis of these oral traditions reveals two major themes in how multiple Suns are conceptualised. The first theme is that the people describe the existence of multiple Suns, typically as siblings or mother/children, that were once visible in the sky, but their combined heat caused drought or deadly conditions. The oral traditions from Yirrkalla and Oenpelli feature this theme. The tradition recorded from Elcho Island simply describes the existence of multiple Suns in the past, but makes no mention of them having too much combined heat. Another tradition from Yirrkalla also mentions that sometimes two Suns are visible. The second theme describes multiple Suns that move across the sky at different times of the year. These traditions are found in Tiwi, Worrorra, and Cape Bedford traditions. These traditions describe the Suns walking different pathways across the sky at different times of the year.

The traditions in Theme #1 describe the appearance of multiple Suns in the sky at the same time. The traditions stipulate that the Sun-woman had children who wished to join her on the journey across the sky, but were sent back to the Underworld since their combined heat would kill everyone. The stories from Yirrkalla and Oenpelli in Arnhem Land, and Cape Bedford, Queensland state that the multiple Suns in the sky at some stage created conditions that were too hot. Those in Arnhem Land suggest that the Suns are still occasionally visible on rare occasions, but indicate that this does not last for long. For example, the traditions from Yirrkalla indicate the Sun is accompanied by one or two smaller Suns that are visible at dawn, but the smaller Suns disappear as the Mother Sun rises in the sky. At night the Sun journeys across the Underworld, where she can rest and hide. These traditions suggest an observation of a specific atmospheric phenomenon and an explanation as to the reason for this phenomenon, assigning it meaning.

The traditions in Theme #2 do not state that the Suns are in the sky at the same time. Rather, these seem to be explanations of the gradual change in the angle of the ecliptic throughout the year, from winter solstice to summer solstice. The traditions say the Suns take different pathways between 'summer' and 'winter'. Ethnographic fieldwork conducted by the first author (Hamacher) with Meriam elders on Mer (Murray Island) in the eastern Torres Strait supports this. Elder Alo Tapim describes two pathways the Sun takes, on in the dry season and one in the wet season. The focus there is more on the angle of the Sun in the sky – how it is higher during the Kuki (wet season from December to March) and lower in the Sager (dry season from April to November).

All of the traditions contain symbolic and social meaning, but are those that describe multiple Suns in the sky at the same time consistent with observable natural phenomena that closely match the depictions in the traditions describes in Theme #1? If so, and if the phenomena could be seen from the geographic locations of the stories, this could show how Aboriginal people observe rare atmospheric phenomena. Of course, this does not mean that the description is *not also* symbolic in nature. Many astronomical observations by Aboriginal people include both witnessed natural phenomena and allegorical tales featuring symbolic meaning (Haynes, 1990; Haynes, 2000), although many Aboriginal cultures would not necessarily differentiate the two.

There are few natural phenomena that could explain the appearance of multiple Suns in the sky. Some might consider a supernova, but one required to rival the Sun in brightness would need to occur within 100 light years of Earth. This could cause significant damage to our planet's biosphere, potentially leading to a global extinction event (Clark et al, 1977; Whitten et al., 1976). No evidence for this has been discovered to date. Additionally, a supernova would remain in the sky for a period of weeks to months as it gradually faded from view. This is not consistent with the description in the tradition.

In her book, "*The Seven Sisters of the Pleiades: Stories from around the World*", Bardi woman and author Munya Andrews (2004: 91) proposed that the Aboriginal traditions "speak of a time when the Sun was once a binary star", proposing that this may have been the star Sirius (Alpha Canis Majoris), challenging scientists to explore this hypothesis (*ibid*: 92). Sirius is itself a binary star system. Sirius A is an A-class star visible to the naked eye. It has twice the mass and 25.4 times the luminosity of our Sun, with an age of approximately 225-250 million years (Croswell, 2005). Sirius B, the smaller companion star (not visible to the naked eye), is a white dwarf that formed around 120 million years ago. Prior to becoming a white dwarf, Sirius B was more massive than Sirius A. These two stars have a highly elliptical orbit, with a separation distance of 8.2 AUs at perihelion and 31.5 AUs at aphelion. As a G-class star, our Sun is currently 4.6 billion years old with a lifespan of 10 billion years, making it nearly 20 times the age of the Sirius system. The proper motion of Sirius (which is actually moving toward the Earth with a radial velocity of –5.5 km/s) conflicts with this explanation. Finally, a stellar companion ejected from the Solar System is inconsistent with the low eccentricity orbits of the Earth and other planets (Benacquista, 2012).

Although the presence of a distant physical companion to the Sun has been proposed by astronomers since the 1980s (Whitmire and Jackson, 1984), no evidence for a stellar dwarf companion in our Solar System has ever been found, despite systematic infrared telescopic surveys capable of detecting small dwarf stars within 10 light years of the Sun (Kirkpatrick et al., 2011). Mamajek *et al*. (2015) calculated that a small binary red/brown dwarf system, collectively called Scholz's star, passed through the outer Oort Cloud (0.8 light years from the Sun) around 70,000 years ago. Although the star system's visual magnitude ($v_{mag}$ = 11.3) was well below the threshold of human visibility ($v_{mag}$ ~ 6.5), one of the stars is magnetically active and occasionally flares. It could have been just visible for minutes to hours during flare-ups. But even then, it was too faint to come close to rivalling the Sun (or Moon, or Venus) in brightness.

The presence of two (or more) Suns in the sky could be explained by an atmospheric effect known as *parhelia* (from Greek meaning 'beside the Sun'). This is a type of halo created when sunlight is refracted through hexagonal light crystals in cirrus or cirrostratus clouds high in the atmosphere, where the temperature is very low. The ice crystals act as prisms, refracting the sunlight. The hexagonal shape of these crystals when oriented randomly can form a 22° halo (or other arcs) around the Sun or Moon (Figure 3A). If the ice crystals sink through the atmosphere, they tend to align vertically, causing the sunlight to refract horizontally. This can result in concentrated regions of light at points 22° horizontal to the Sun on either side, appearing as smaller 'mock Suns' (Figure 3B). Varying atmospheric conditions can result in multiple haloes, arcs, and multiple (three or more) mock Suns, depending on the orientation of the ice crystals. Other related phenomena are 44° arcs and haloes, parry arcs, crepuscular arcs, upper tangent arcs, circum-zenithal arcs, and parhelic circles (Figure 3C).

Solar arcs and rays are described in Aboriginal traditions across Victoria. For example, Moporr elder Weerat Kuyuut and his family told Dawson (1881: 101) that an:

> *"upper crepuscular arch in the east at sunset is called* Kuurokeheear puuron, *meaning "white cockatoo twilight" and the under arch is* Kappiheear puuron, *meaning "black cockatoo twilight." Crepuscular rays in the west after sunset are called "rushes of the Sun", while those before sunrise are called* Kullat *meaning "peep of day."*

Parhelia are relatively common and can be seen anywhere in the world, but are more frequent in colder regions, such as the Arctic, Antarctic, and winter skies over high or low latitude locations. For example, Weerat Kuyuut and his family explained that the halo arches come from the constellation Orion (Dawson, 1881: 101), which sets at dusk in June during the Australian winter. Although less common in warmer climates and tropical latitudes, solar haloes and parhelia can still be visible when local conditions are right. The first author witnessed multiple haloes (but no parhelia) high in the sky over Sydney on 13 February 2014, when the ground temperature was 29° C (Figure 3A).

Parhelia are most commonly seen when the Sun has a low altitude, particularly at dawn when the local temperature is the lowest. Mock suns can appear on either side of the Sun and tend to disappear as the Sun rises. The mock suns generally appear to be much smaller than the Sun itself, but historical records report rare instances of two Suns of similar size appearing high in the sky. A well-known example is the famous twin-Suns seen over Rome at noon on 22 February 1929 for nearly 30 minutes (Anonymous, 1929). Reports say the two suns were the same size and appeared to be linked by a luminous arch.

The sighting of two smaller suns accompanying the Sun at dawn is consistent with observations of parhelia. The Yolngu traditions from Yirrkalla and Oenpelli noted that the smaller, accompanying suns disappeared as the Sun rose into the sky, attributing this to the Sun woman sending them back to the underworld so their combined heat does not cause death or drought. It is also consistent with the second tradition from Yirrkalla that says on occasion, two suns are visible in the sky.

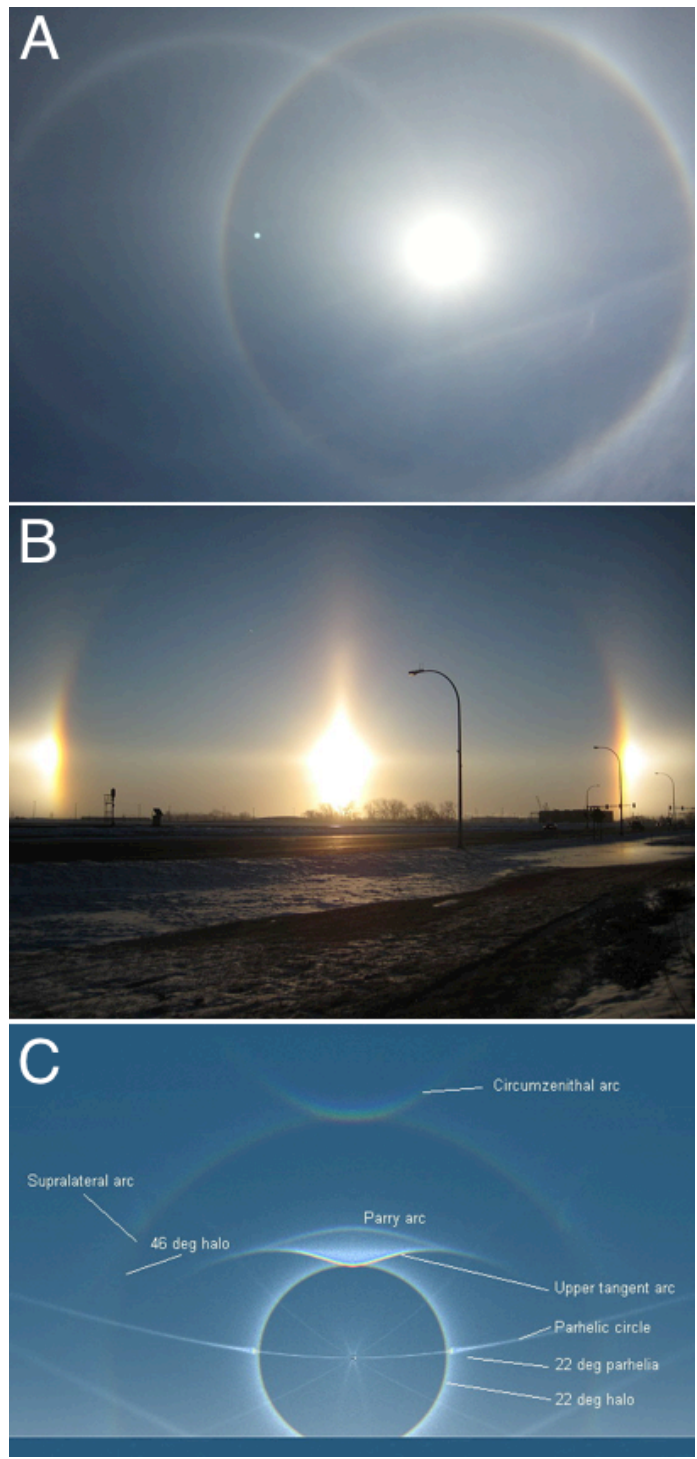

**Figure 3:** (A) Haloes around the Sun as seen over Sydney in February 2014. Photo by D.W. Hamacher. (B): Parhelia photographed from Fargo, North Dakota, USA in February 2008. Photo reproduced from Wikipedia Commons. (C) Various types of halo and arc phenomena-caused by ice crystals in the atmosphere. Screenshot from HaloSim3 software

The description of parhelia in Aboriginal oral traditions from Arnhem Land is consistent with cultural traditions around the world, which often portray the Sun as having twin children. For example, in Mesoamerica, the Lancandon Maya have six oral traditions about two Suns in the sky (Kováč, 2017). One involves animosity between the two suns, which lead to the elimination of one of them. This particular tradition could be traced back to an earlier Mesoamerican traditions in which two suns existed since time began, and one had to be purged. Only a smattering of the stories speak about the occurrence of two suns being visible at the same time in the sky, but they seem to be symbolic in nature and not necessarily descriptions of observed parhelia. However, a tradition from the Teotihuacan people of Mexico might suggest observations of parhelia. It describes twin Suns that existed during the creation of the world, in which the duplicate sun was eliminated by the other sun while it was climbing up a tree or a tower (ibid.: 19), similar to how parhelia tend to disappear as the Sun climbs into the sky.

In Arctic and sub-Arctic regions, parhelia are well known to First Nations communities. An Alaskan tradition speaks about a woman named Katxu'n and her daughter (Olson, 1967: 44). Many animal-men approached the mother for the daughter's hand in marriage, but she declined the offer. One day, the Sun-man approached the mother and spoke about his qualification: "Each morning I lift myself up in the east. Everyone is glad. I make things dry. I make people warm." The mother consented to the marriage and the daughter bore four sons. One day, the Sun found out his wife was being unfaithful to him and became angry. He took his four sons to the sky and it is said that on a clear day, the sons are visible in the sky as parhelia.

## 5    CONCLUDING REMARKS

We examine Aboriginal Australian oral traditions describing multiple Suns in the sky. We find two major themes present in these accounts: one that describes multiple Suns in the sky at dawn that do not all rise together for fear their combined heat would burn the Earth. The second theme relates to two Suns that travel along different pathways at different times of the year. We argue that the first theme is a narrative account of parhelia, while the second theme describes the change in the angle of the ecliptic throughout the year, with extremes near the solstices.

Parhelia fit the descriptions in oral tradition and are visible from anywhere on Earth. The oral traditions from Arnhem Land clearly describe two or three companion Suns accompanying the Sun at sunrise. But, for reasons discussed in cultural terms, the companion Sun(s) does not continue to rise, but rather disappear. This shows that Aboriginal people observe and note all forms of astronomical and atmospheric phenomena, assigning them meaning and social purpose.

## 6    ACKNOWLEDGEMENTS

We acknowledge the Aboriginal communities discussed in this paper and recognise their traditional knowledge and intellectual property. We also thank Gillian Stewart for bringing the subject to our attention. Hamacher is funded by Australian Research Council project DE140101600 and conducted ethnographic fieldwork in the Torres Strait under Human Research Ethics approval no. HC15035.